\documentclass[a4paper,usenatbib]{mnras}

\usepackage{color}
\usepackage{graphicx}
\usepackage{amssymb}
\usepackage{amsmath}
\usepackage{aas_macros}

\newcommand{\de}{\ensuremath{{\rm d}}}
\newcommand{\be}{\begin{equation}}
\newcommand{\ee}{\end{equation}}
\newcommand{\bear}{\begin{eqnarray}}
\newcommand{\ear}{\end{eqnarray}}
\newcommand{\nline}{\nonumber \\}
\newcommand{\f}{\frac}

\newcommand{\Msun}{\mbox{M}_{\odot}}

\newcommand{\eqn}[1]{equation~\eqref{#1}}

\newcommand{\fig}[1]{Figure~\ref{#1}}
\newcommand{\figs}[1]{Figures~\ref{#1}}

\begin{document} 
\title[Epoch of reionization using fractal dimensions]
{Studying neutral hydrogen structures during the epoch of reionization using fractal dimensions}
\author[Bandyopadhyay, Choudhury \& Seshadri]
{Bidisha Bandyopadhyay$^{1}$\thanks{Email: bidisharia@gmail.com}, T. Roy Choudhury$^{2}$\thanks{Email: tirth@ncra.tifr.res.in}, T. R. Seshadri$^1$\thanks{Email: seshadri.tr@gmail.com}\\
$^1$Department of Physics and Astrophysics, University of Delhi, Delhi 110007, India\\
$^2$National Centre for Radio Astrophysics, Tata Institute of Fundamental Research, Ganeshkhind, Pune 411007, India}

\maketitle
\date{\today}

\begin{abstract}
Fractal dimensions can be used to characterize the clustering and lacunarities in density distributions. We use generalized fractal dimensions to study the neutral hydrogen distribution (HI) during the epoch of reionization. Using a semi-numeric model of ionized bubbles to generate the HI field, we calculate the fractal dimensions for length scales $\sim 10 h^{-1} {\rm cMpc}$. We find that the HI field displays significant multifractal behaviour and is not consistent with homogeneity at these scales when the mass averaged neutral fraction $\bar{x}_{\rm HI}^M \gtrsim 0.5$. This multifractal nature is driven entirely by the shapes and distribution of the ionized regions. The sensitivity of the fractal dimension to the neutral fraction implies that it can be used for constraining reionization history. We find that the fractal dimension is relatively less sensitive to the value of the minimum mass of ionizing haloes when it is in the range $\sim 10^9 - 10^{10} h^{-1} \Msun$. Interestingly, the fractal dimension is very different when the reionization proceeds inside-out compared to when it is outside-in. Thus the multifractal nature of HI density field at high redshifts can be used to study the nature of reionization.
\end{abstract}


\section{Introduction}

The study of the neutral hydrogen (HI) 21~cm signal from the reionization epoch at $6 \lesssim z \lesssim 15$ is one of the most important observational probes to study the structure formation in early times \citep{2001PhR...349..125B,2006astro.ph..3149C,2006PhR...433..181F,2012RPPh...75h6901P}. Low frequency radio telescopes like the Low Frequency Array (LOFAR)\footnote{\tt http://www.lofar.org/} \citep{2013A&A...556A...2V}, the Precision Array for Probing the Epoch of Reionization (PAPER)\footnote{\tt http://eor.berkeley.edu/} \citep{2014ApJ...788..106P}, the Murchison Widefield Array (MWA)\footnote{\tt http://www.mwatelescope.org/} \citep{2011AAS...21813206B}, and the Giant Metrewave Radio Telescope (GMRT)\footnote{\tt http://www.gmrt.tifr.res.in} \citep{2011MNRAS.413.1174P,2013MNRAS.433..639P} have already taken data to detect the statistical fluctuations in the 21~cm signal and are precursors for the upcoming experiments like the Square Kilometre Array (SKA)\footnote{\tt http://www.skatelescope.org/} and Hydrogen Epoch of Reionization Array (HERA)\footnote{\tt http://reionization.org/}.

The first generation of these experiments are aimed at detecting the signal of reionization by measuring the HI power spectrum \citep{2013MNRAS.433..639P,2014PhRvD..89b3002D,2015ApJ...809...61A}, or equivalently the two-point correlation function, which is believed to contain a significant amount of information on the distribution of the HI. However, almost all models of reionization predict that the signal is highly non-Gaussian \citep{2006MNRAS.372..679M,2007ApJ...659..865L,2007MNRAS.379.1647W}, mainly because it is determined by the shape and growth of the ionized bubbles forming around the ionizing sources (galaxies). In that case, it is possible to extract additional information on the reionization process from the higher order statistics and other related diagnostics. There have been attempts to carry out studies using, e.g., one point statistics like the skewness \citep{2015MNRAS.451..467S,2015MNRAS.454.1416W}, higher order correlations like the the bispectrum \citep{2005MNRAS.358..968B,2006MNRAS.366..213S,2007ApJ...662....1P,2015MNRAS.451..266Y,2016MNRAS.458.3003S} and the trispectrum \citep{2005MNRAS.363.1049C} and Minkowski functionals \citep{2006MNRAS.370.1329G,2008ApJ...675....8L,2014JKAS...47...49H,2016arXiv160202351Y}.

A different way of characterizing the underlying structure of the HI field beyond the two-point functions is to use the ``fractal dimension'' \citep{Mandelbrot,Martinez}. One of the main features of the fractal dimension is that it directly provides information about the deviation from a homogeneous distribution \citep{Feder, Seshadri1, Seshadri2}. It also enables us to know about the degree of clumping or lacunarity which are the tracers of the nonlinear processes. The method has been applied extensively to the galaxy surveys \citep[see, e.g.,][]{1987PhyA..144..257P,1998PhR...293...61S,1999A&A...351..405B,2001ApJ...554L...5M,2010MNRAS.405.2009Y}. Using galaxy surveys \citet{1992PhyA..185...45C} showed that the galaxy structures show self-similar behaviour up to very large scales. Subsequently \citet{1990MNRAS.242..517M,1995PhR...251....1B} showed that such nature is exhibited only up to a certain finite scale. \citet{1999A&A...351..405B} have shown that not only fractal patterns are observed in the galaxy surveys but also the galaxy correlations exhibit a multifractal behaviour upto a scale $\sim 100~h^{-1}$ Mpc and become homogeneous at larger scales. Using SDSS data, \citet{2005MNRAS.364..601Y,2009MNRAS.399L.128S} concluded that the galaxy distribution is homogeneous on scales larger than $60-70 h^{-1}$ Mpc.

In this work, we apply the method of fractal dimensions to study the HI distribution from the reionization epoch. In particular, our main aim is to verify if the method is sensitive to the fraction of mass that is neutral at a given redshift. In that case, the method can be used as a diagnostic tool to determine the degree of ionization of the medium using observations. Our study is mainly based on simulating the HI distribution using an excursion set based semi-numeric model on the lines of \citet{2007ApJ...654...12Z,2009MNRAS.394..960C}, and estimating the fractal dimension for different values of the parameters involved.

The structure of the paper is as follows: In Section \ref{sec:fractal} we discuss the basic formalism for calculating the fractal dimensions for a density field. The semi-numeric models for generating the HI field during reionization is described in Section \ref{sec:sem-num}. The main results of our work are presented in Section \ref{sec:results}. We summarize our conclusions in Section \ref{sec:discussion}. In this work, we use the flat $\Lambda$CDM cosmological model with parameters given by $\Omega_m = 0.305, \Omega_b = 0.048, h = 0.679, n_s = 0.96, \sigma_8 = 0.827$ \citep{2014A&A...571A..16P}.



\section{Fractal dimensions}
\label{sec:fractal}

The fractal dimensions can be used for determining the degree of clustering and lacunarity of the distribution of a set of points in space. These points could be deterministic or statistical and can be characterized by a fractal parameter, also called the fractal dimension, which for a homogeneous and isotropic continuous field takes the value of the ambient dimension of space in which it is embedded. An inhomogeneous distribution for which clustering has a scaling relationship with distance can be described by a fractal dimension which is always different from the ambient dimension.

There are various methods for quantifying fractal dimensions for a set of discrete points. Two such methods that are commonly used are the \emph{box counting} method and the \emph{correlation dimension} method by direct counting \citep[for a detailed discussion see][]{1995PhR...251....1B}. For the purpose of this paper, where we will be dealing mainly with continuous density fields, the second method will turn out to be more useful. We first discuss the calculation of the correlation dimension for a discrete set of points, and later show how to extend the calculations for a continuous density field.

\subsection{Correlation dimension for a set of discrete points}

Let us first discuss the fractal dimensions for the case where the distribution we aim to study consists of $N$ discrete points \citep{1993PhRvE..47.3879B,1995PhR...251....1B}. This could, e.g., be the distribution of galaxies. In this method we first randomly select $M$ of these $N$ points indexed by $i$. Treating each of these points as centres, we draw spheres of increasing radii and count the number of sample points $N_i(<r)$ that are inside a sphere as a function of its radius $r$, i.e., 
\be
N_i(<r)=\sum_{j=1}^{N }\Theta\left(r - |{\mathbf x_i }-{\mathbf x_j}| \right),
\ee
where $\Theta(x)$ is the step function defined as equal to $1$ for $x>0$ and $0$ for $x<0$, and ${\mathbf x}_i$ denotes the position vector of the $i$th point.

In astrophysical context, each such data point is often weighted by a weight factor $w_j$. For example, in galaxy surveys one could miss out some of the galaxies because of their faintness and the weight factor is a way to compensate for it.  This gives an effective number $N_{i,{\rm eff}}(<r)$ defined as the sum of all the weights of points that are inside a sphere (of radius $r$) and centred at the point $i$
\be
N_{i,{\rm eff}}(<r)=\sum_{j=1}^{N }w_j\Theta(r-\mid {\mathbf x_i }-{\mathbf x_j}\mid).
\ee

The integrated correlation function $C_2(r)$ of the given distribution can then be defined as
\begin{equation}
C_2(r)=\frac{1}{N M}\sum_{i=1}^M N_{i,{\rm eff}}(<r). \label{C_2_def}
\end{equation}
Since $C_2$ is the integrated two-point correlation function, one could ask if it is possible to construct quantities that reflect the higher order $n$-point correlation functions. This is indeed possible by generalizing \eqn{C_2_def} to the $q$th moment of the distribution, i.e.,
\begin{equation}
C_q(r)=\frac{1}{N M}\sum_{i=1}^M (N_{i,{\rm eff}}(<r))^{q-1}.
\end{equation}
In the case of a scaling behaviour, $C_q(r)$ can be expressed as a power law as $C_{q} \propto r^{D_q (q-1)}$ for the description under consideration. Then the generalized dimension corresponding to the $q$th moment is given by
\begin{equation}
 D_{q}=\frac{1}{q-1}\frac{\de~\ln~C_{q}(r)}{\de~\ln~r}.
\label{eq:Dq}
\end{equation}
This quantity is also known as the Minkowski-Bouligand dimension or simply the fractal dimension. A monofractal has a constant $D_q$ irrespective of $q$ while for multifractals $D_q$ varies with $q$. If a distribution is a monofractal and has the fractal dimension as its ambient dimension, then that distribution is homogeneous. In case of a homogeneous (space filling) distribution in three dimensions, the value of $D_q$ should be 3.

\subsection{Correlation dimension for a continuous distribution}

The above prescription, however, is not suited for applying to HI maps which are usually represented as continuous field defined at points on a uniform grid. Let us assume that the density field $\rho_i$ is defined on $N_{\rm grid}$ grid points, where $i$ runs from 1 to $N_{\rm grid}$. We can then generalize the definition of $C_q(r)$ to 
\be
C_q(r) = A_q \sum_{i=1}^{N_{\rm grid}} m_i~\left[m_i(<r)\right]^{q-1},
\ee
where $m_i$ is the mass contained in the $i$th grid cell, $m_i(<r)$ is the mass contained within a sphere of radius $r$ centred around the $i$th cell and the sum is over all grid points. The quantity $A_q$ is a proportionality constant whose exact value is not important for our analysis. Note that the term $m_i$ in the expression ensures that cells with more mass carry more weight as they are likely to be sampled more frequently (as one would have done while working with discrete points).

It is obvious that $m_i = \rho_i \times V_{\rm cell}$, and $m_i(<r) = \rho_i(<r) \times 4 \pi r^3 / 3$, where $V_{\rm cell}$ is the volume of each grid cell (which is constant for a uniform grid) and $\rho_i(<r)$ is the density within a sphere of radius $r$ around the $i$th grid. Instead of the density field, it is often more convenient to work in terms of the overdensity 
$\Delta_i \equiv \rho_i / \langle \rho \rangle$, where $\langle \rho \rangle$ is the globally-averaged value of the density field. In that case, we can write
\be
C_q(r) = A_q \sum_{i=1}^{N_{\rm grid}} \Delta_i~\left[\Delta_i(<r)~r^3\right]^{q-1},
\label{eq:Cqr}
\ee
where the unimportant constants have been absorbed within the definition of $A_q$. The generalized dimension $D_q$ can be obtained simply using \eqn{eq:Dq}. Note that for a homogeneous distribution we expect $\Delta_i$ to be constant and hence $C_q(r) \propto r^{3 (q-1)}$, thus leading to $D_q = 3$, which is consistent with our expectations.

Interestingly, the quantity $C_2(r)$ can be related to the usual two-point correlation function $\xi(r)$, and can be shown to have the value
\bear
C_2(r) &=& \f{1}{V} \int_V \de^3 r'~[1 + \xi(r')]~\Theta(r - r')
\nline
&=& \f{4 \pi}{V} \int_0^r \de r'~r'^2~[1 + \xi(r')].
\ear
In terms of the averaged correlation function $\bar{\xi}(r)$, the above expression becomes
\be
C_2(r) = \f{4\pi}{3 V} r^3 \left[1 + \bar{\xi}(r)\right].
\ee

The $C_q(r)$ for $q \neq 2$ can be related to the higher order correlation functions of the distribution being considered. In fact, it can be shown that $C_q(r)$ for $q \geq 2$ is a combination of a series of correlation functions starting from the $q$-point correlation function to the two-point correlation function \citep{1995PhR...251....1B}. Thus the quantities $C_q$ and $D_q$, in principle, contain information on the existing non-Gaussianity in the underlying field. One can also see that $D_q$ for positive values of $q$ is mostly determined by the high density cells, while that for negative $q$ gets more weight from underdense points. In the case of neutral hydrogen distribution, the positive $q$'s will trace the distribution of dense neutral regions, while the negative $q$s would trace the distribution of ionized or partially ionized regions. Of course, for a density field which is homogeneous at scales of interest, one expects $D_q = 3$, i.e., equal to the ambient dimension of the density field.

\section{The HI density field}
\label{sec:sem-num}

In this section, we briefly outline the method for generating the HI density field which has been used for our study. The method is based on (i) obtaining the dark matter density field at redshifts of interest by running a $N$-body simulation, the publicly available GADGET-2\footnote{\tt http://wwwmpa.mpa-garching.mpg.de/gadget/} \citep{2005MNRAS.364.1105S} in our case, (ii) locating the positions of the haloes using the friends-of-friends (FoF) algorithm and (iii) using an excursion set based semi-numeric technique to identify the ionized regions. 

The length of the $N$-body simulation box used in this study is $100 h^{-1}$cMpc and it contains $1024^3$ collisionless dark matter particles. Assuming that a FoF halo should contain at least 20 particles, the mass of the smallest halo turns out to be $1.6 \times 10^9 h^{-1}$M$_{\odot}$. This mass is somewhat higher than the minimum mass of star forming haloes which cool via atomic transitions, however, it is similar to the typical minimum mass threshold set by the radiative feedback \citep{2008MNRAS.384.1525S}.

The ionization field is obtained using a condition based on the excursion set formalism \citep{2007ApJ...654...12Z,2009MNRAS.394..960C} where, for a given grid cell $i$, one constructs spheres of different radii $r$ around the cell and calculates the the collapsed fraction $f_{i, \rm coll}(<r)$ within the sphere. The grid cell $i$ is flagged as ionized if the condition 
\be
\zeta f_{i, \rm coll}(<r) \geq 1,
\ee
is satisfied for any values of $r$, where $\zeta$ is the ionizing efficiency. In case the condition is not satisfied for all $r$, we flag the cell as partially ionized with a ionization fraction $\zeta f_{i, \rm coll}$, where $f_{i, \rm coll}$ is the collapsed fraction within the $i$th cell \citep{2008MNRAS.386.1683G}. The process is repeated for all cells in the box and thus the ionization field is created. We set the value of $\zeta$ at the redshift of interest by demanding that the resulting HI maps have the required value of the mass averaged neutral fraction $\bar{x}_{\rm HI}^M$. The above method of identifying the ionized cells is performed in a relatively coarser grid of resolution $0.5 h^{-1}$~cMpc, implying $N_{\rm grid} = 200^3$ in the simulation box we are using.

The quantity of our interest is the mass averaged neutral fraction $\Delta_{\rm HI} \equiv x_{\rm HI} \Delta_b$ evaluated at each grid cell, where $x_{\rm HI}$ is the neutral fraction and $\Delta_b$ is the baryon overdensity. The field $\Delta_{\rm HI}$ measures the fluctuations in the HI density. In 21~cm observations, the quantity that is directly measured is the differential brightness temperature \citep{2012RPPh...75h6901P}
\begin{equation}
T_b =  \bar{T}_b~x_{\rm HI}~\Delta_b~\frac{T_{\rm spin}- T_{\rm CMB}}{T_{\rm spin}},
\end{equation}
where $\bar{T}_b \approx 22$mK $[(1+z) / 7]^{1/2}$ is the mean temperature \citep{2009MNRAS.394..960C}, $T_{\rm spin}$ is the HI spin temperature and $T_{\rm CMB}$ is the CMB radiation temperature. As we can see, $T_b$ is directly proportional to $\Delta_{\rm HI}$ when $T_{\rm spin} \gg T_{\rm CMB}$, a condition that is expected to hold once the IGM becomes more than $\sim 10$ per cent ionized \citep{2015MNRAS.447.1806G}.

\section{Results}
\label{sec:results}

We shall present the main findings of our analysis in this section. The main quantity of interest would be $D_q$ which is expected to be equal to 3 for a homogeneous distribution. The values of $q$ studied in this work lie between $-5$ and $+5$. For values of $q$ smaller than $-5$, we find the $D_q$ to be highly fluctuating with large statistical errors, while for $q > +5$, the $D_q$ does not seem to contain any significant features which may lead to new information. To calculate $D_q$, we essentially need to compute the logarithmic slope of $C_q(r)$. We compute this in the range of length scales given by $10 \leq r / (h^{-1} {\rm cMpc}) \leq 15$. The lower limit of $r$ is set by the fact that $C_q(r)$ does not exhibit power-law behaviour for scales $r \lesssim 10 h^{-1}~{\rm cMpc}$, while the upper limit corresponds to the typical largest scales for which we can reliably estimate the correlations in our simulation box without being affected by the limited box size. These scales are quite appropriate for the first generation low-frequency experiment aiming to detect the reionization signal \citep{2013MNRAS.433..639P,2014PhRvD..89b3002D,2015ApJ...809...61A}. The $D_q$ is obtained from $C_q(r)$ by fitting a power-law and estimating the index using the least-square method.

We have validated our method by applying to a set of random Poisson distributed points in a cubical box and calculating the fractal dimensions. We found that the fractal dimension is consistent with the ambient dimension $D_q = 3$ for all values of $q$.

 \begin{figure*}
   \includegraphics[width=0.95\textwidth]{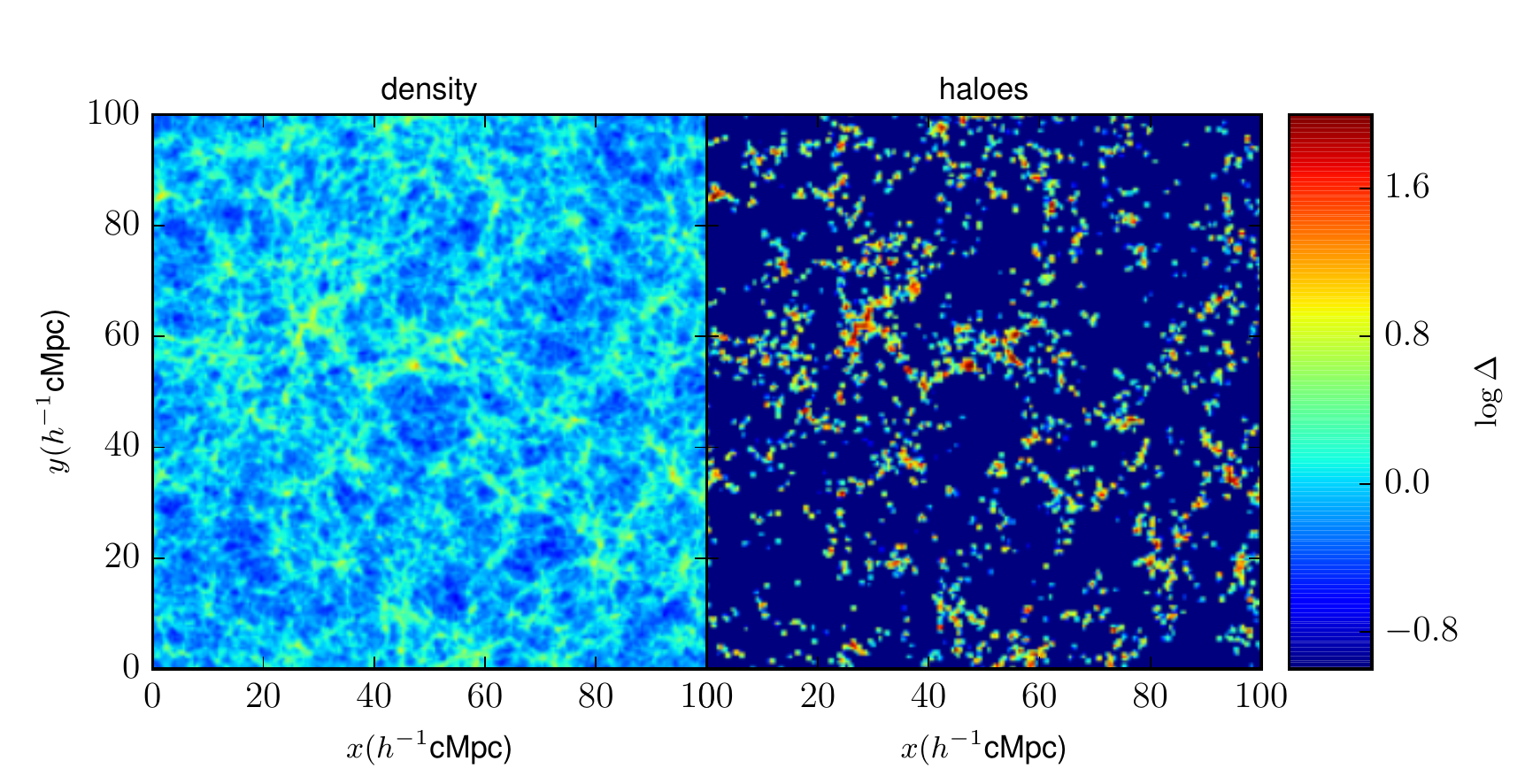}
   \caption{A $0.5 h^{-1}$cMpc thick slice of the smoothed density field (left) and the halo field (right) from our simulation box at $z = 7$.} 
   \label{fig:density_halo}
 \end{figure*}

 \begin{figure}
   \includegraphics[width=0.45\textwidth]{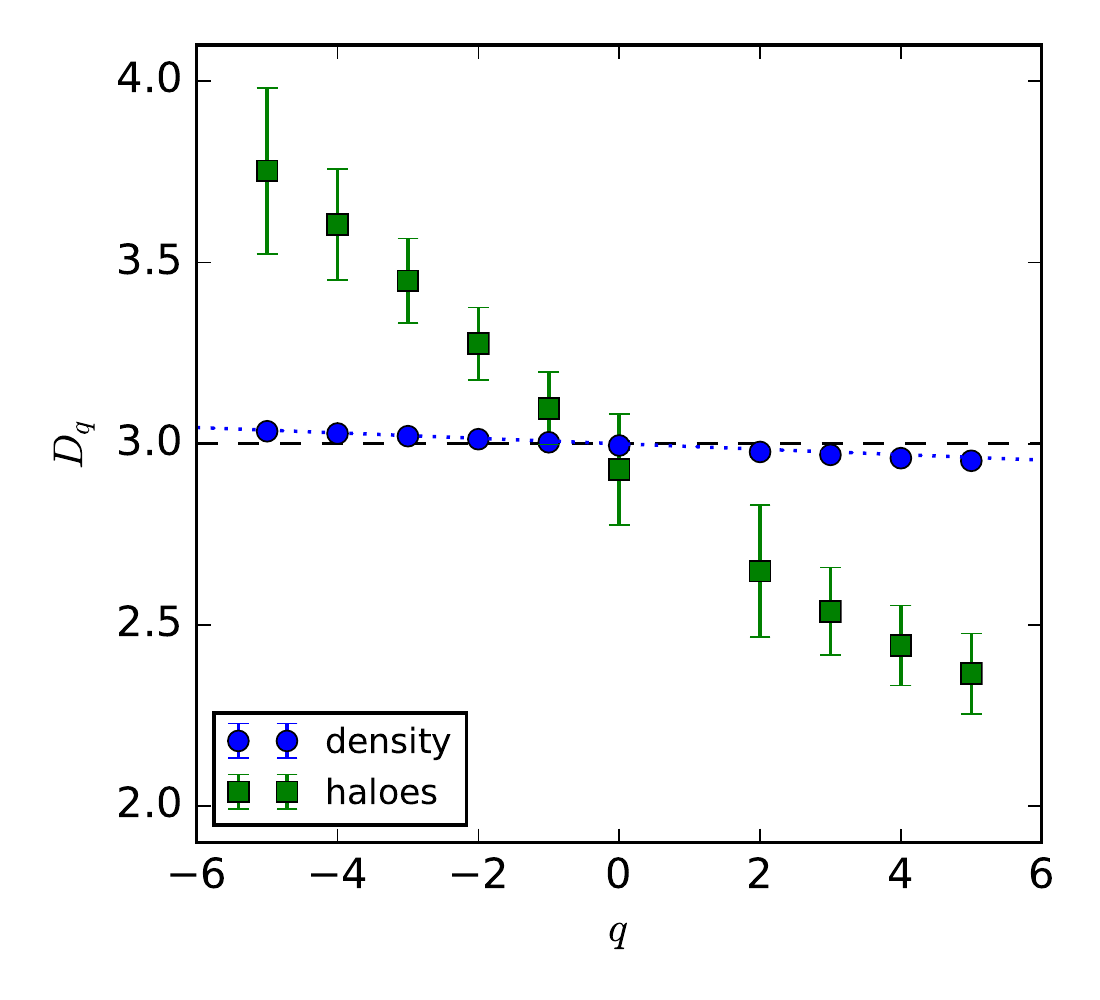}
   \caption{The fractal dimension $D_q$ for the cosmological density field (circles) and the halo field (squares) at $z = 7$. The error-bars are obtained from 20 random sub-samples of points in the simulation box, see the text for more details. The error-bars for the cosmological density field are smaller than the size of the circles. The dotted lines represent the function $D_q - 3 \propto -q$, normalized so as to match the $D_q$ corresponding to the density field.} 
   \label{fig:Dq_density}
 \end{figure}

\subsection{The density and the halo fields}

 \begin{figure*}
   \includegraphics[width=0.95\textwidth]{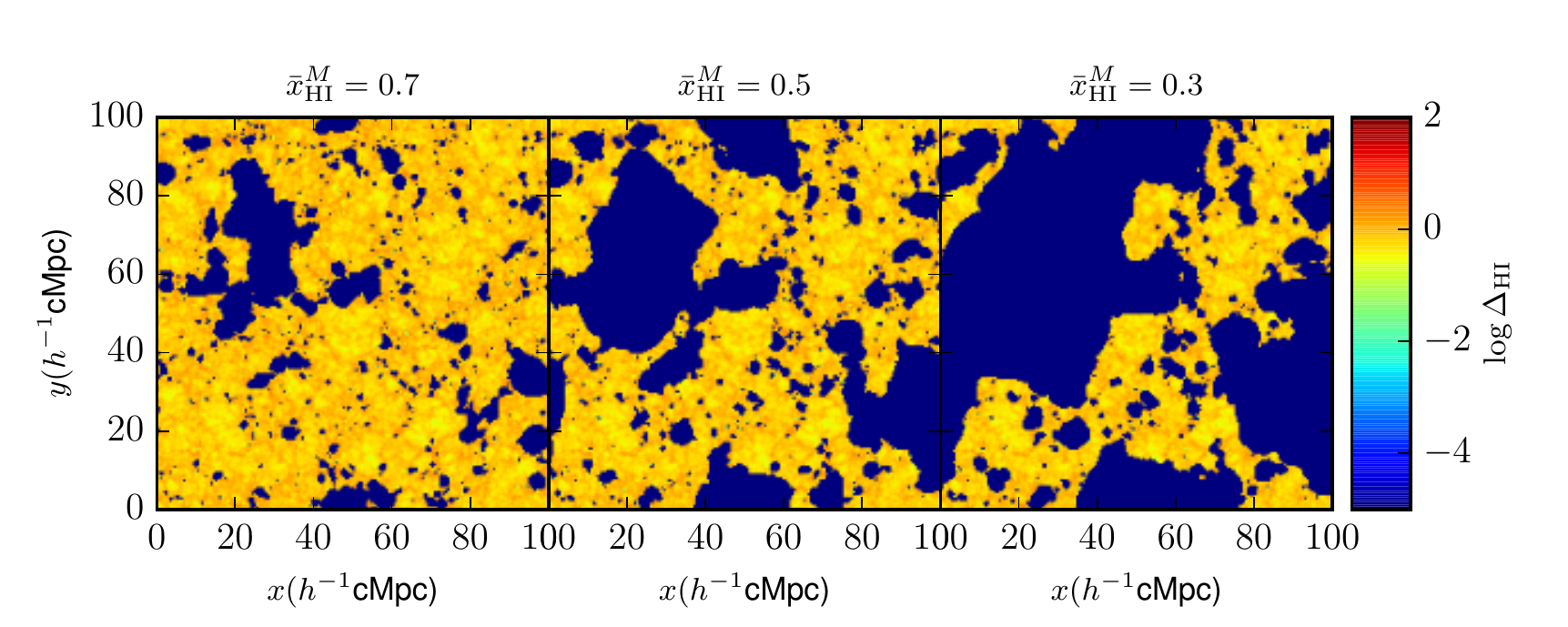}
   \caption{A $0.5 h^{-1}$cMpc thick slice of the HI density field $\Delta_{\rm HI}$ for $\bar{x}_{\rm HI}^M = 0.7$ (left) and $0.3$ (right) from our simulation box at $z = 7$.} 
   \label{fig:HI}
 \end{figure*}

 \begin{figure}
   \includegraphics[width=0.45\textwidth]{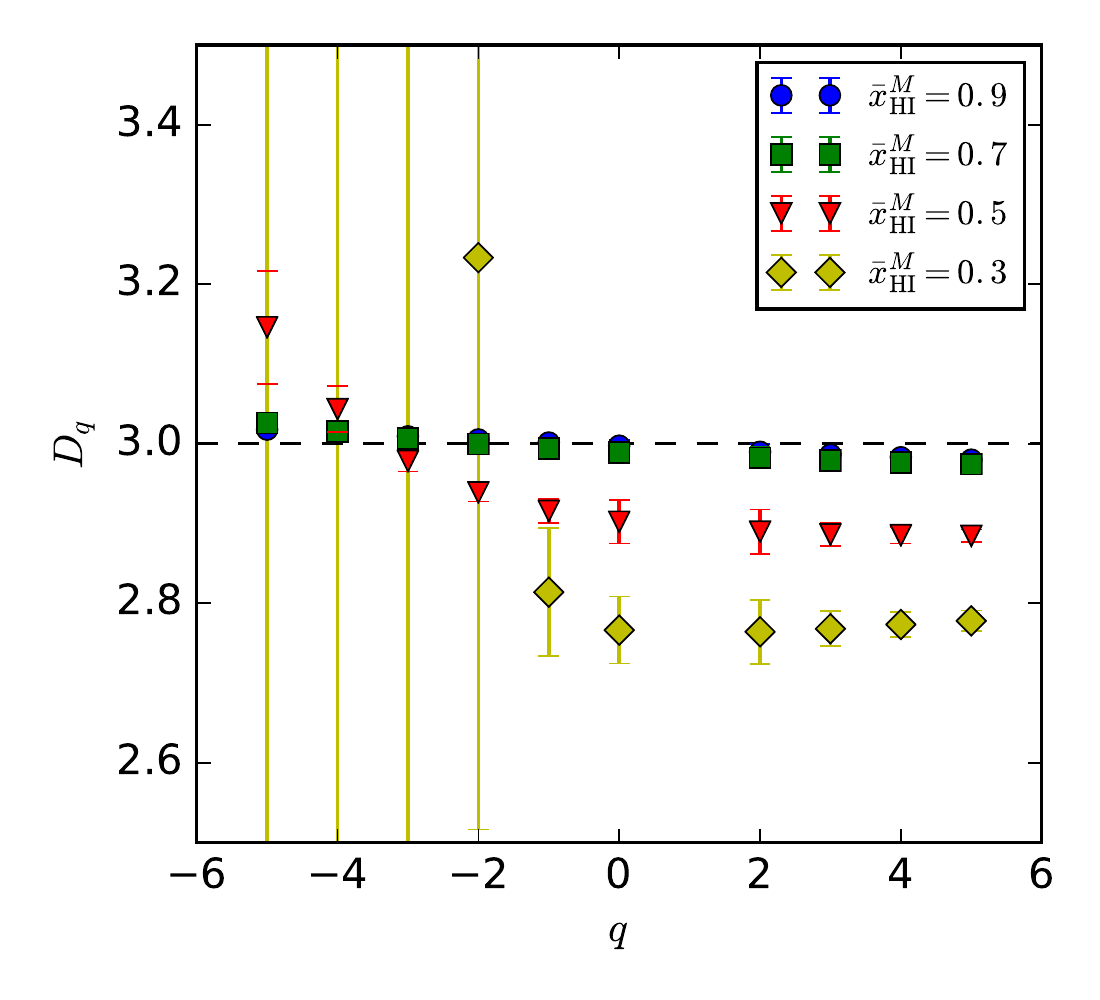}
   \caption{The fractal dimension $D_q$ for the HI density field $\Delta_{\rm HI}$ at $z = 7$ for different values of $\bar{x}_{\rm HI}^M$. The error-bars are obtained from 20 random sub-samples of points in the simulation box, see the text for more details. The points for $\bar{x}_{\rm HI}^M = 0.9$ and $\bar{x}_{\rm HI}^M = 0.7$ are practically indistinguishable.} 
   \label{fig:Dq_HI}
 \end{figure}

Next let us compute the fractal dimensions for the underlying cosmological density field and the field of the dark matter haloes which are contributing to the ionizing photons (i.e., the haloes identified through the FoF algorithm that satisfy the minimum number of particles criterion).  The density field is obtained by smoothing the particle positions in the simulation box over a $200^3$ grid, while the halo field is obtained from the corresponding halo positions (weighed appropriately by their masses). The halo field thus is a smooth representation of the underlying galaxy distribution weighed by their ionizing emissivities. The two-dimensional maps of the two fields are shown in \fig{fig:density_halo} where we have plotted a slice of thickness $0.5 h^{-1}$cMpc from our simulation box at $z = 7$. It is obvious that the halo field is significantly more clustered and less homogeneous than the density field. The resulting $D_q$ as well as the statistical errors for the two fields at $z = 7$ are shown in \fig{fig:Dq_density}.

 \begin{figure}
   \includegraphics[width=0.45\textwidth]{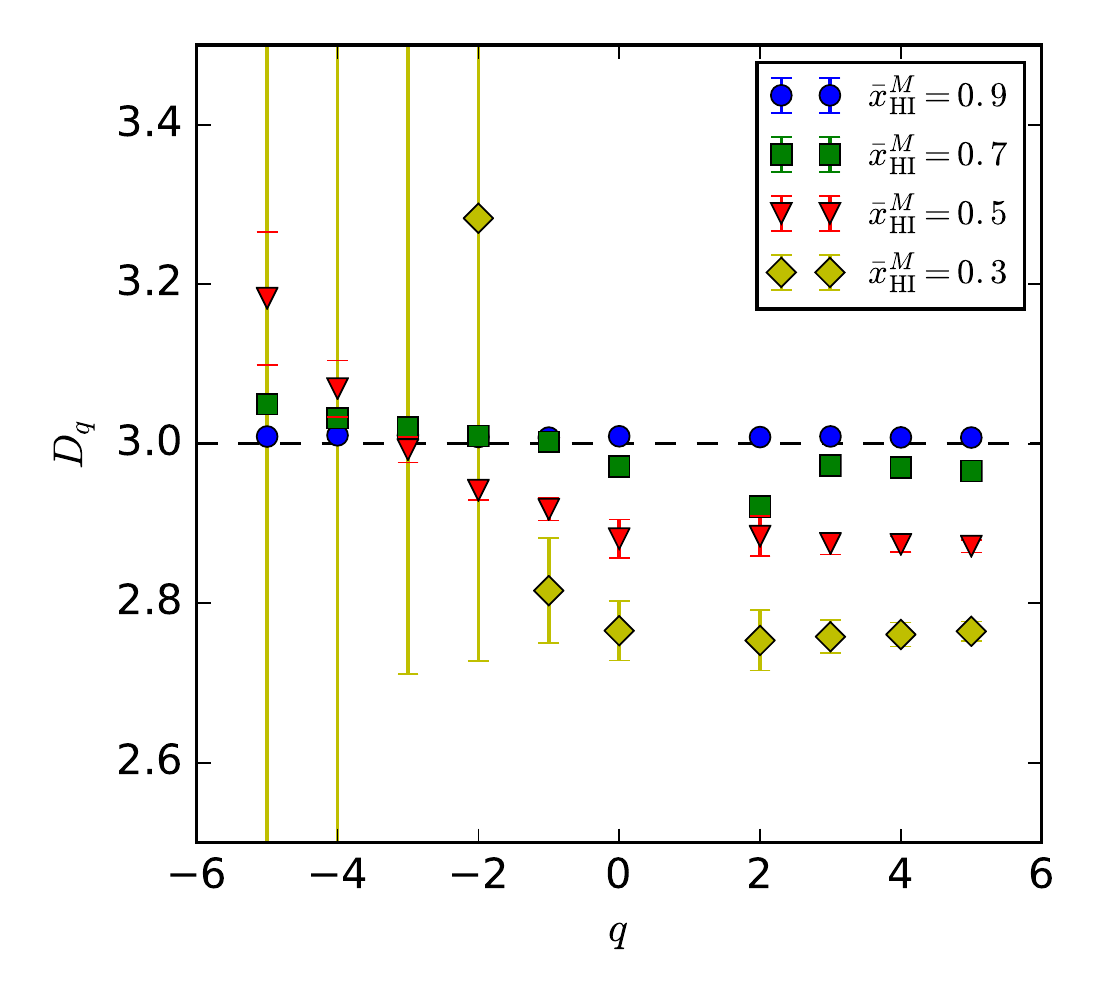}
   \caption{The fractal dimension $D_q$ for the neutral fraction field $x_{\rm HI}$ at $z = 7$ for different values of $\bar{x}_{\rm HI}^M$. The error-bars are obtained from 20 random sub-samples of points in the simulation box, see the text for more details. Note that we plot the results for the values $\bar{x}_{\rm HI}^M$ that would be obtained for the true HI density field $\Delta_{\rm HI}$. In other words, the respective curves in \fig{fig:Dq_HI} and this one have the same value of the efficiency $\zeta$.} 
   \label{fig:Dq_xHI}
 \end{figure}

 \begin{figure*}
   \includegraphics[width=0.9\textwidth]{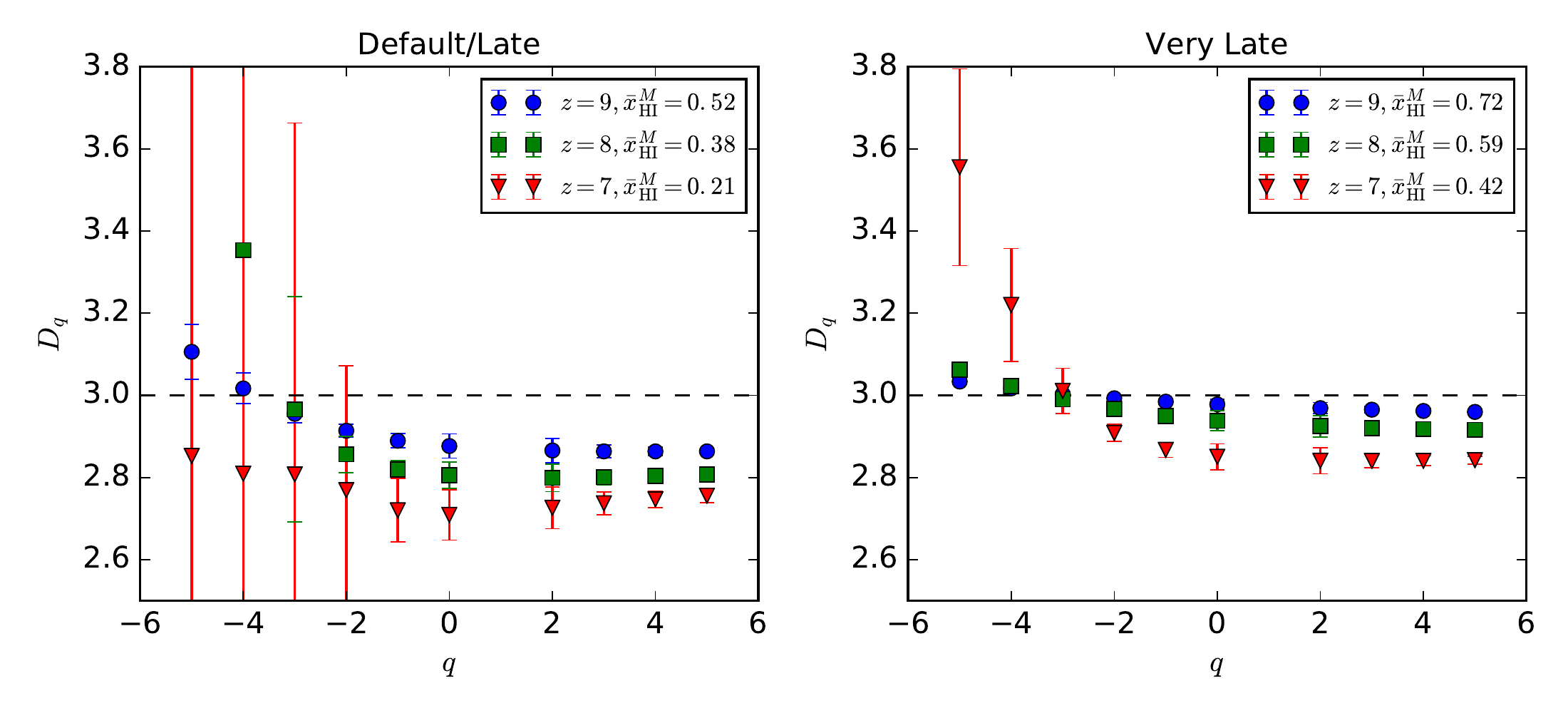}
   \caption{The fractal dimension $D_q$ for the HI density field $\Delta_{\rm HI}$ at different redshifts for two reionization histories. The left hand panel is for the Default/Late model with $\tau_{\rm el} = 0.068$, while the right hand panel is for the Very Late model with $\tau_{\rm el} = 0.055$.} 
   \label{fig:Dq_reion}
 \end{figure*}

The estimation of statistical errors on $D_q$ ideally requires multiple realizations of the density and halo fields which in turn requires us to run the $N$-body simulations multiple times using different initial conditions. This is somewhat expensive, hence we calculate the errors using a relatively simpler method. We randomly divide the $N_{\rm grid}$ cells into 20 independent groups and calculate $C_q(r)$ using \eqn{eq:Cqr} for each group. We then estimate the errors on $C_q(r)$ from these subsamples, which subsequently translates into errors on $D_q$. 

We can see from \fig{fig:Dq_density} that the values of $D_q$ for the cosmological density field are very close to 3 for all $q$, though they are not consistent with 3 within error-bars particularly for large values of $q$. Interestingly, we find that the results for the density field are consistent with $D_q - 3 \propto -q$. As was shown by \citet{2008MNRAS.390..829B}, the departure of $D_q$ from the ambient dimension (to the first order) in the regime of weak clustering $q \bar{\xi} \ll 1$ is given by
\be
D_q - 3 \approx -\f{3 q}{2} \left | \f{\de \bar{\xi}}{\de \ln r} \right |.
\ee
The behaviour of $D_q$ for the density field is thus consistent with the above expression, which is not surprising given the fact that the clustering of the density field at scales $\sim 10 h^{-1}$~cMpc at $z \sim 7$ is expected to be weak.
 
The halo field, on the other hand, shows $D_q$ that is markedly different from the ambient dimension of 3, implying that the distribution of haloes are not homogeneous at the scales of $\sim 10-15 h^{-1}~{\rm cMpc}$. This is what is expected from the clustering of galaxies which are biased compared to the underlying density field. We checked and found that the value of $D_q$ does approach 3 for the halo field if we restrict ourselves to only larger scales $\sim 50 h^{-1} {\rm cMpc}$, however, the error-bars too increase in that case because of less number of modes sampled and the results are less reliable because the length scales probed become comparable to the size of the simulation box.

\subsection{The HI density field}

 \begin{figure*}
   \includegraphics[width=0.95\textwidth]{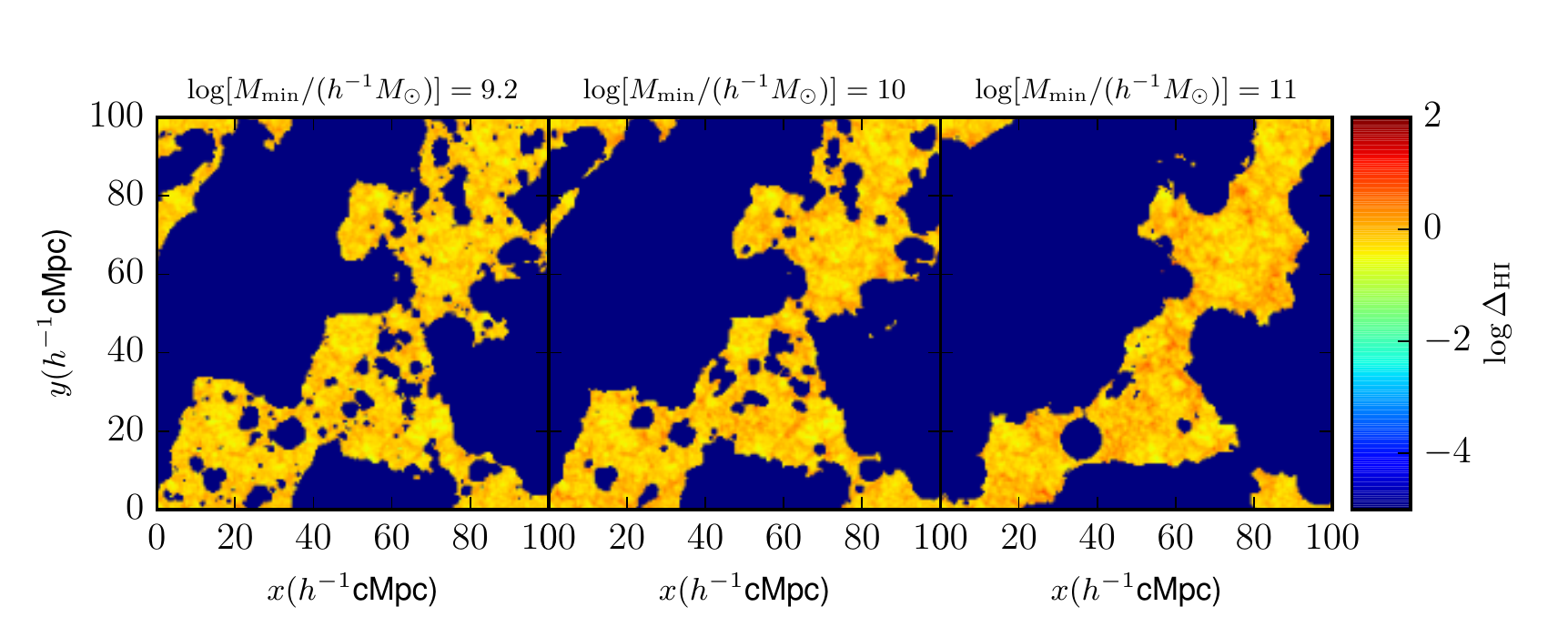}
   \caption{A $0.5 h^{-1}$cMpc thick slice of the HI density field $\Delta_{\rm HI}$ for $M_{\rm min} = 1.6 \times 10^9 h^{-1} \Msun$ (left), $10^{10} h^{-1} \Msun$ (middle) and $10^{11} h^{-1} \Msun$ (right) from our simulation box at $z = 7$. All the three cases are normalized such that $\bar{x}_{\rm HI}^M = 0.3$.} 
   \label{fig:Mmin}
 \end{figure*}

 \begin{figure}
   \includegraphics[width=0.45\textwidth]{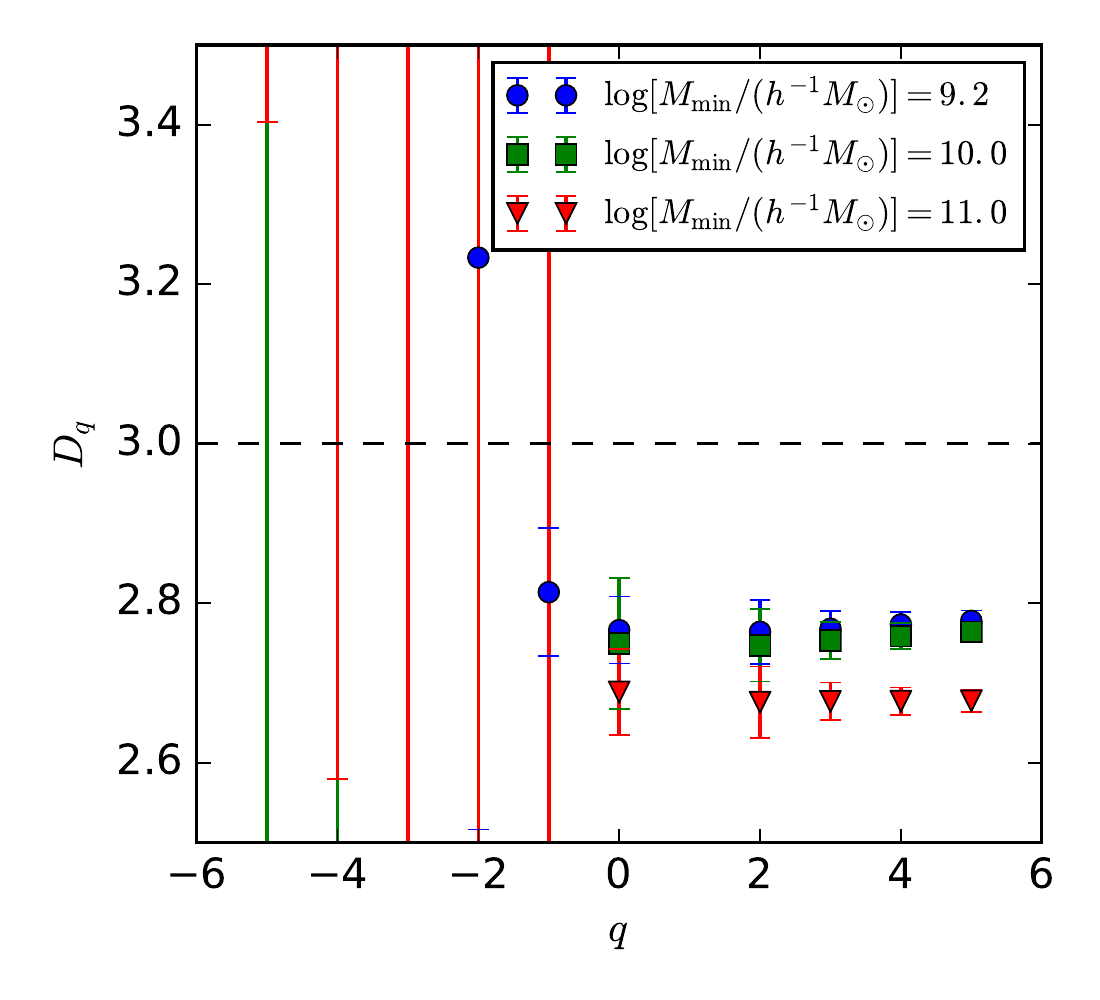}
   \caption{The fractal dimension $D_q$ for the HI density field $\Delta_{\rm HI}$ at $z = 7$ for three different values of $M_{\rm min}$. All the models have been normalized to give $\bar{x}_{\rm HI}^M = 0.3$.} 
   \label{fig:Dq_Mmin}
 \end{figure}

 \begin{figure*}
   \includegraphics[width=0.95\textwidth]{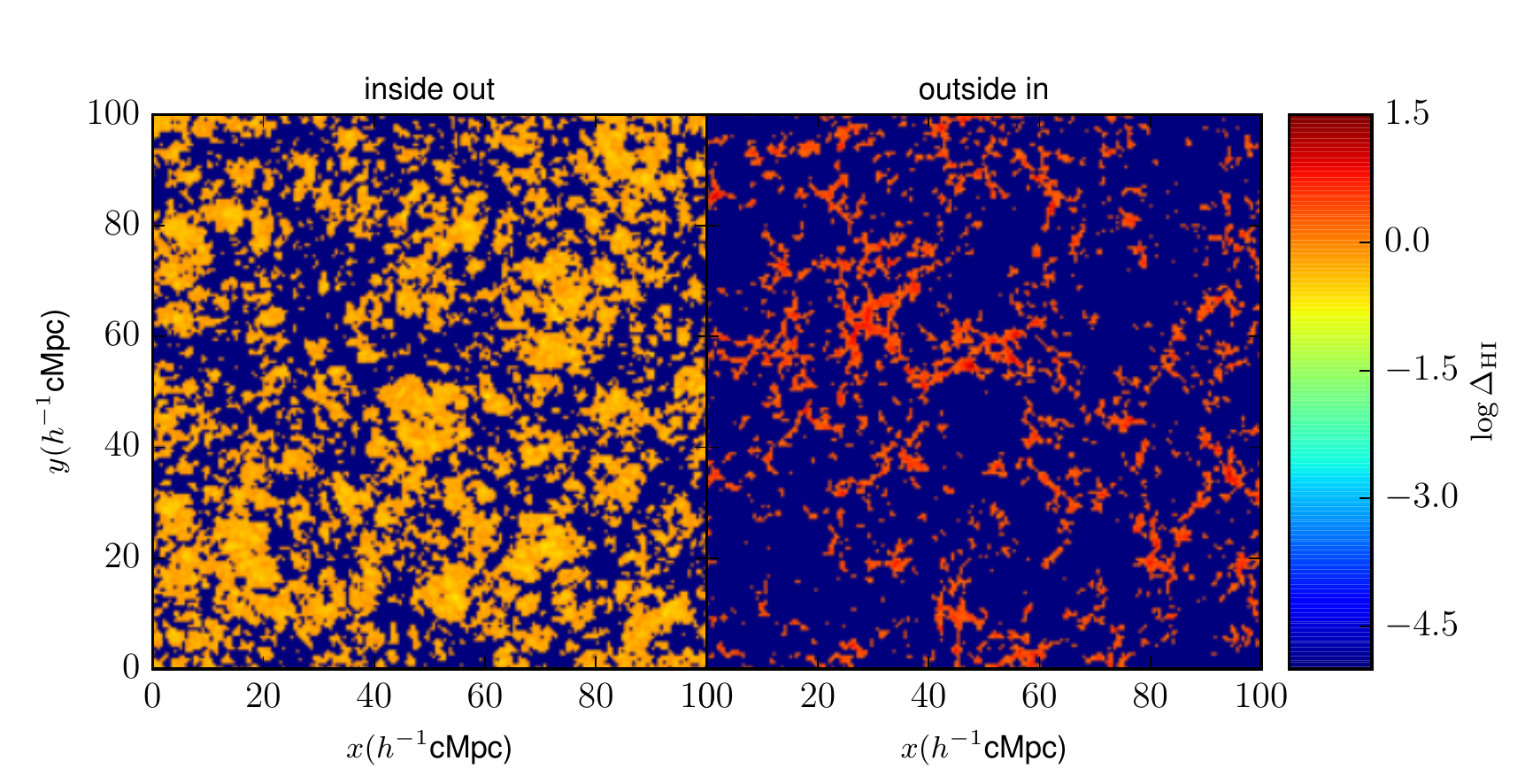}
   \caption{A $0.5 h^{-1}$cMpc thick slice of the HI density field $\Delta_{\rm HI}$ for two extreme reionization models, i.e., one which is strictly inside-out (left) and strictly outside-in (right) at $z = 7$. Both the cases are normalized such that $\bar{x}_{\rm HI}^M = 0.3$.} 
   \label{fig:density_threshold}
 \end{figure*}

 \begin{figure*}
   \includegraphics[width=0.9\textwidth]{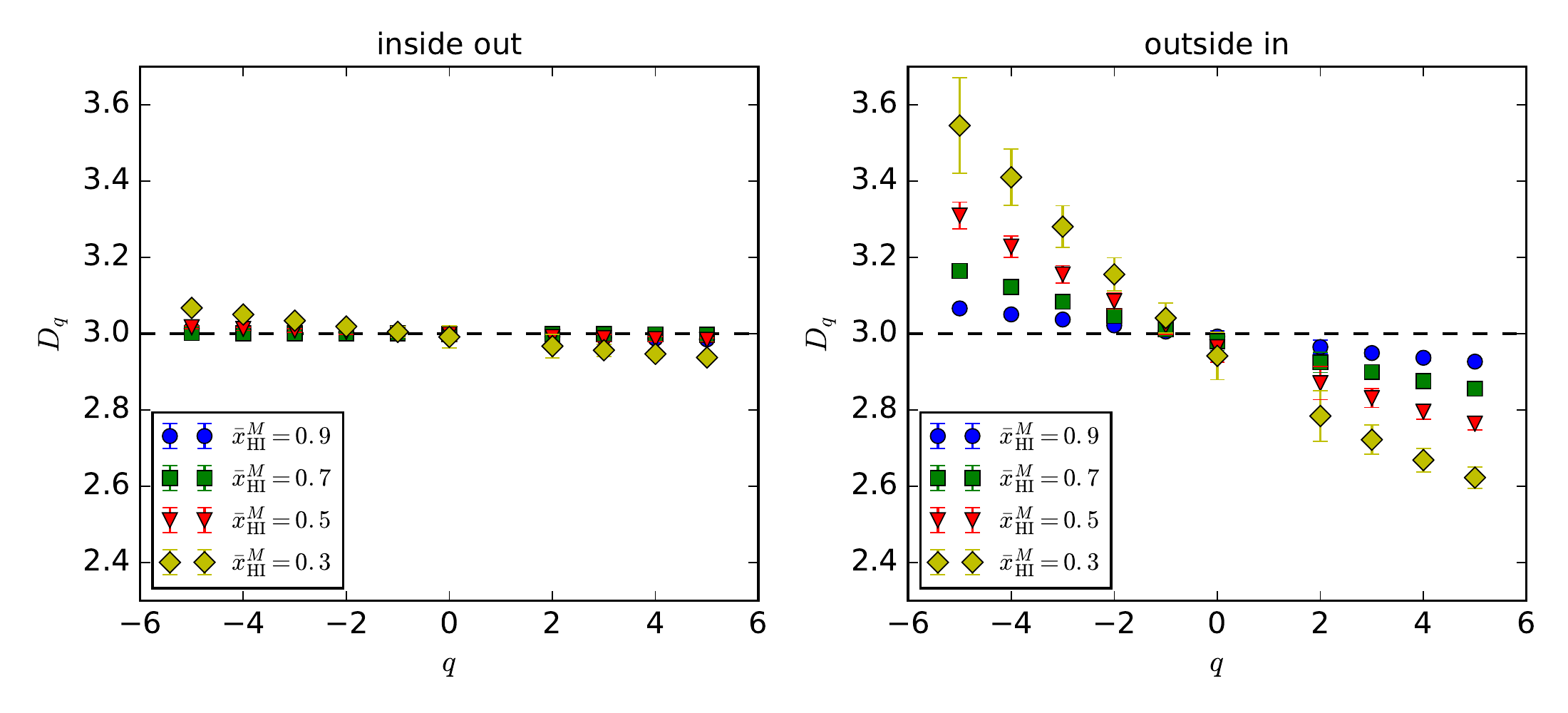}
   \caption{The fractal dimension $D_q$ for the HI density field $\Delta_{\rm HI}$ at $z = 7$ for the strictly inside-out (left) and outside-in (right) models for different values of $\bar{x}_{\rm HI}^M$.} 
   \label{fig:Dq_density_threshold}
 \end{figure*}

We next calculate the fractal dimensions for the HI density field $\Delta_{\rm HI}$ obtained by the semi-numeric method described in Section \ref{sec:sem-num}. Let us first inspect the two-dimensional maps for three values of the mass averaged neutral fractions $\bar{x}_{\rm HI}^M = 0.7, 0.5$ and $0.3$ at $z = 7$ in \fig{fig:HI}. The dark blue regions represent the ionized regions which can be seen to have percolated significantly when $\bar{x}_{\rm HI}^M = 0.3$ (right hand panel). One can also see that for high values of $\bar{x}_{\rm HI}^M$ (left hand panel) the field looks relatively more homogeneous. Hence we expect $D_q$ to be close to 3 at early stages of reionization and deviate from 3 as reionization progresses with time.

The plots of $D_q$ for $\bar{x}_{\rm HI}^M = 0.3, 0.5, 0.7, 0.9$ at $z = 7$ are shown in \fig{fig:Dq_HI}. Note that we vary the value of the efficiency parameter $\zeta$ thus obtaining different values of $\bar{x}_{\rm HI}^M$ at the fixed redshift, whereas in reality the $\bar{x}_{\rm HI}^M$ would evolve with $z$. We do this to disentangle any effect arising from the evolution of the density field from those arising because of the evolution of the ionized regions. Later, we will discuss the results for a realistic reionization history. The error-bars are calculated using the method described in the previous subsection. While interpreting the error-bars, one should keep in mind that the volume spanned by typical 21~cm observations of the reionization epoch will be significantly larger than our simulation box, and the large number of modes probed in the scales of interest should help in decreasing the statistical errors. In case the statistical errors are small, the possibility of distinguishing the $D_q$ from the ambient dimension of 3 would depend on the observational noise. We defer the discussion of such errors to a future work and concentrate mainly on understanding the properties of the fractal dimensions of the HI density field.

The first point to note is that the error-bars on $D_q$ blow up for $q \leq -2$ for $\bar{x}_{\rm HI}^M \lesssim 0.5$. This is because the number of points contributing to the sum in the expression for $C_q(r)$ becomes smaller as the medium gets more ionized. In addition, for negative values of $q$, the sum will be dominated by only a few points having relatively small $\Delta_{\rm HI}$, thus leading to large statistical fluctuations. In fact, we found that the error-bars for $x_{\rm HI} = 0.1$ blow up even for $q$ as large as $-1$.

One can see that for high values of the neutral fraction $\bar{x}_{\rm HI}^M \gtrsim 0.7$, the fractal dimension $D_q$ is almost identical to that of the underlying density field. The values of $D_q$ of all $q$ are close to 3, though not consistent with 3 based on the error-bars we have calculated. The value of $D_q$ starts deviating from 3 as the value of $x_{\rm HI}$ decreases, and hence the HI distribution shows prominent multifractal nature. We find that for $x_{\rm HI} = 0.5 (0.3)$, the value of $D_q$ for $q \geq 2$ is $\sim 2.9 (2.8)$, thus showing that the HI distribution is not homogeneous at length scales as large as $\sim 10-15 h^{-1}~{\rm cMpc}$. In principle the measurement of $D_q$ at these values of $q$ can be one diagnostic of the mass averaged neutral fraction. In particular, significantly large departures from the value of 3 would imply relatively larger values of the ionization fraction.

To gain some insight into why this deviation of $D_q$ from the ambient dimension of 3 arises, we plot the fractal dimension for only the ionization fraction field $x_{\rm HI}$. In this case, the fluctuations in the HI density field sourced by the underlying density field $\Delta_b$ is absent. The results are shown in \fig{fig:Dq_xHI}. Note that we plot the results for the values $\bar{x}_{\rm HI}^M$ that would be obtained for the true HI density field. In other words, the respective curves in \figs{fig:Dq_HI} and \ref{fig:Dq_xHI} have the same value of the efficiency $\zeta$. It is clear that the $D_q$ is practically the same for the $\Delta_{\rm HI}$ and $x_{\rm HI}$ fields, thus implying that the multifractal nature of the HI field is essentially determined the distribution of the ionized regions. This essentially follows from the fact that the density field is almost homogeneous at these length scales.

We now show the evolution of $D_q$ over redshifts for realistic reionization histories. We choose two reionization models motivated and consistent with the measurements of the electron scattering optical depth $\tau_{\rm el}$. The first model corresponds to a reionization history giving $\tau_{\rm el} = 0.068$ and is similar to the Default/Late model of \citet{2016MNRAS.463.2583K}. The model is motivated by the measurements of $\tau_{\rm el}$ obtained using the Planck temperature and lensing data \citep{2016A&A...594A..13P} and consistent with constraints obtained by \citet{2015MNRAS.454L..76M,2016MNRAS.455.4295G}. We also consider another model where reionization begins somewhat late and proceeds relatively faster, giving $\tau_{\rm el} = 0.055$. This second model is similar to the Very Late model of \citet{2016MNRAS.463.2583K}. The value of $\tau_{\rm el}$ is similar to that obtained by combining the Planck CMB temperature anisotropy data with the low-multipole polarization data \citep{2016arXiv160503507P}.

The resulting $D_q$ for the HI density field $\Delta_{\rm HI}$ for different redshifts are shown in \fig{fig:Dq_reion}. As expected, since in the Default/Late model the neutral fractions are less at a given redshift, the $D_q$ shows prominent departures from homogeneity. In contrast, the Very Late model is considerably neutral even at $z = 7$, thus the deviations from 3 are relatively less.

The above models have been obtained assuming the minimum mass of the star-forming haloes to be $M_{\rm min} = 1.6 \times 10^9 h^{-1} \Msun$, same as the least massive halo in our simulation box. However, there is still not much understanding as to what should be the values of $M_{\rm min}$ at high redshifts. In absence of any negative radiative feedback, the gas within haloes as small as $\sim 10^6 \Msun$ can cool via molecular hydrogen transitions and form stars. On the other hand, destruction of molecules by Lyman-Werner radiation can increase the $M_{\rm min}$ to $\sim 10^8 \Msun$ where the gas can cool via atomic transitions. Another limit on $M_{\rm min}$ arises because of the radiation feedback from reionization which can prohibit star formation in haloes $\lesssim 10^9 \Msun$ in the photoheated regions \citep{1998MNRAS.296...44G,2008MNRAS.384.1525S,2016ApJ...825L..17G}. In addition, feedback from supernovae too can expel gas from the low-mass galaxies thus suppressing star formation \citep{2003ApJ...599...38B,2013MNRAS.428.2966P,2013MNRAS.428.2741W}. Uncertainties on $M_{\rm min}$ can also arise because of our limited knowledge of the escape of ionizing photons from the host halo. While some simulations claim that the escape of photons are relatively easier for high mass $\gtrsim 10^{10} \Msun$ haloes \citep{2008ApJ...672..765G}, whereas others have argued that escape is most efficient in small mass $\sim 10^7 - 10^8 \Msun$ haloes \citep{2010ApJ...710.1239R,2014MNRAS.442.2560W,2014ApJ...788..121K}.

In order to understand how the multifractal nature of the HI density field is affected by $M_{\rm min}$, we compute $D_q$ for different values of the minimum mass. While doing this comparison, we fix the redshift to be $z = 7$ and ensure that all the cases have the same value of the mass averaged neutral fraction, chosen to be $\bar{x}_{\rm HI}^M = 0.3$. The maps of the HI density field for three values of $M_{\rm min} = 1.6 \times 10^9 h^{-1} \Msun, 10^{10} h^{-1} \Msun$ and $10^{11} h^{-1} \Msun$ are shown in \fig{fig:Mmin}. Clearly, there is some similarity in the overall structure of the HI field which follows from the fact that the haloes broadly trace the peaks of the underlying density field. We see that for smaller values of $M_{\rm min}$ (left hand panel) there exist numerous ionized regions of smaller sizes that are not present for the higher $M_{\rm min}$.  For the extreme value of $M_{\rm min} = 10^{11} h^{-1} \Msun$, we see that the neutral regions are more concentrated and have larger sizes.

The behaviour of $D_q$ for different $M_{\rm min}$ is shown in \fig{fig:Dq_Mmin}. Clearly, the curves look quite similar in their shape, possibly because the broad structure of the HI density fields for the three cases are similar. We also find that there is hardly any difference in the results for $M_{\rm min} = 1.6 \times 10^9 h^{-1} \Msun$ and $10^{10} h^{-1} \Msun$, thus indicating that $D_q$ is relatively insensitive to the value of $M_{\rm min}$. It might thus be possible to constrain the value of $x_{\rm HI}$ from the measurements of $D_q$ if $M_{\rm min}$ lies in this range. However we also find that there is some significant difference in $D_q$ when $M_{\rm min}$ takes a extremely high value of $10^{11} h^{-1} \Msun$. The deviation of $D_q$ from the ambient value of 3 is more for higher values of $M_{\rm min}$. This is mainly because the HI density field is more clustered for higher $M_{\rm min}$ as we have seen from the maps in \fig{fig:Mmin}.

Finally, we investigate how the fractal dimensions behave when we modify the nature of the reionization process. By default, the excursion set based reionization models like the one used in this work are ``inside-out'' at large scales, in the sense that high density regions are ionized first followed by the low density voids \citep{2004ApJ...613....1F,2006MNRAS.369.1625I,2007ApJ...669..663M,2011MNRAS.411..955M}. This is because the ionizing sources are preferentially formed at large scale density peaks. Of course, this does not mean that the process is strictly inside-out at all scales, for example, a low density cell in the simulation box that is nearer to a source is more likely to be ionized earlier than a high density cell that is far away for any source. However, there is an additional complication which does not conform to this simple picture which arises from the ``sinks'' of ionizing photons \citep{2000ApJ...530....1M}. These are essentially the dense regions capable of self-shielding the radiation and thus remaining neutral. The presence of these sinks can make the reionization ``outside-in'' particularly at small scales \citep{2009MNRAS.394..960C}. Accounting for these regions require either very high dynamic range simulations \citep{2015MNRAS.446..566M,2015MNRAS.452..261C}, or additional conditions for determining the ionized cells \citep{2005MNRAS.363.1031F,2009MNRAS.394..960C,2014MNRAS.440.1662S}.

Rather than attempting to model the sinks (which can be quite expensive in terms of the computational requirements), we consider two extreme (and somewhat unrealistic) models, one where the reionization is strictly inside-out, while the other where it is strictly outside-in and study the resulting $D_q$. These models are constructed by simply choosing a density threshold, and identifying all regions with density higher (lower) than this as ionized in the inside-out (outside-in) model \citep{2013MNRAS.435..460J}. We choose this threshold to obtain the desired $\bar{x}_{\rm HI}^M$. The maps for the two cases when $\bar{x}_{\rm HI}^M = 0.3$ are shown in \fig{fig:density_threshold}. It is obvious that the topology of reionization is completely different in the two cases. In the inside-out model (left), the neutral regions are somewhat diffuse as they trace the low density voids. The situation is exactly the opposite in the outside-in model (right) where the neutral regions are quite concentrated.

The fractal dimensions for these two models for a different $\bar{x}_{\rm HI}^M$ are shown in \fig{fig:Dq_density_threshold}. The first point to note is that the values of $D_q$ are much closer to 3 for the inside-out model than the corresponding outside-in ones, thus implying that the HI distribution in the inside-out models are relatively more homogeneous at scales $\sim 10 h^{-1}$~cMpc. The reason is that the dense and clustered cells are ionized first in the inside-out models, thus reducing the contrast in the HI field. On the other hand, the HI field in the outside-in models essentially trace the high density regions thus giving rise to less homogeneous structures. In fact, the shape of the $D_q$ curves approach that for the halo field as $\bar{x}_{\rm HI}^M$ decreases. It is thus interesting that the fractal dimensions are quite sensitive to the nature of reionization, and thus could possibly distinguish between various scenarios. One should thus keep in mind that inclusion of photon sinks in the semi-numeric reionization models used in this paper can affect the behaviour of $D_q$. Note that the values of $D_q$ for a reionization model with accurate treatment of the sinks are expected to be bound by the values obtained for these two extreme models, as long as the underlying cosmological density field remains the same.

\section{Discussion and Conclusion}
\label{sec:discussion}

The first detection of the 21~cm signal from reionization, expected soon from the low-frequency experiments, is likely to be done by estimating the power spectrum, or equivalently the two-point correlation function, of the HI density field \citep{2013MNRAS.433..639P,2014PhRvD..89b3002D,2015ApJ...809...61A}. However since the distribution of the ionized bubbles during the reionization epoch is expected to lead to a 21~cm signal that is highly non-Gaussian \citep{2007ApJ...659..865L,2007MNRAS.379.1647W}, a significant amount of information would be contained in the higher order correlations of the distribution. In this work, we have explored the generalized correlation dimension (or the Minkowski-Bouligand dimension) $D_q$ as a diagnostic for understanding the nature of the HI distribution. The quantity $D_q$ has been used extensively for characterizing the multifractal nature of the galaxy distribution at low redshifts \citep{1987PhyA..144..257P,1990MNRAS.242..517M,1992PhyA..185...45C,1995PhR...251....1B,1998PhR...293...61S,1999A&A...351..405B,2001ApJ...554L...5M,2005MNRAS.364..601Y,2009MNRAS.399L.128S,2010MNRAS.405.2009Y}. The feature of $D_q$ is that it contains information about different higher order correlation functions and thus can probe the non-Gaussianity in the HI distribution \citep{1995PhR...251....1B}.

We use a semi-numeric simulation to generate the HI density field which is then used for estimating the $D_q$ at scales $\sim 10 h^{-1} {\rm cMpc}$. Our main findings are as follows:

\begin{itemize}

\item The value of $D_q$ for the underlying baryonic density field is almost equal to (though not consistent with, within statistical error-bars) the ambient dimension 3 at our fiducial redshift $z=7$, thus indicating that the density field can be taken to be homogeneous at scales $\sim 10 h^{-1} {\rm cMpc}$. The $D_q$ for the corresponding halo field (or equivalently the distribution of galaxies weighed according to their ionizing emissivities), however, deviates significantly from 3 at these scales.

\item The $D_q$ is almost equal to 3 for the HI density fluctuations $\Delta_{\rm HI}$ field for mass averaged neutral fraction $\bar{x}_{\rm HI}^M \gtrsim 0.7$, thus indicating the distribution to be almost homogeneous when the IGM is significantly neutral. However, as the IGM gets more and more ionized, the $D_q$ becomes different from 3. For $\bar{x}_{\rm HI}^M \sim 0.5 (0.3)$, the value of $D_q$ for $q \geq -1$ is $\sim 2.9 (2.8)$.

\item The statistical errors on $D_q$ blow up for $q < -1$ for $\bar{x}_{\rm HI}^M \lesssim 0.5$, thus making these moments unsuitable for studying the HI field.

\item The $D_q$ for the HI field is driven mainly by the distribution of the ionized regions, thus making them interesting quantities to study the growth and percolation of the ionized regions.

\item The $D_q$ for $q \geq -1$ is sensitive to the globally mass averaged $\bar{x}_{\rm HI}^M$, while its sensitivity to the minimum mass $M_{\rm min}$ (in the range $\sim 10^9 - 10^{10} h^{-1} \Msun$) of ionizing haloes is relatively small. This indicates that the generalized correlation dimension can be used for constraining the global neutral fraction using upcoming observations.

\item The form of the fractal dimension $D_q$ is, in principle, quite sensitive to the nature of reionization, e.g., whether the process is inside-out or outside-in. The presence of dense regions that are sinks of ionizing radiation, which are not included in the analysis here, can affect the value of $D_q$.

\end{itemize}

The study presented in this work needs to be henceforth expanded in order to apply to the observed HI maps. In particular, one needs to calculate how the generalized dimension $D_q$ responds to the noise in the observations. Such calculations would help us infer the amount of observation time in different telescopes that would be required to distinguish between different ionization fractions. Also, it must be noted that the interferometric observations cannot measure the mean value of the field, thus one would have to work with a mean subtracted field $\propto \Delta_{\rm HI} - \bar{x}_{\rm HI}^M$. This would lead to mixing of the different moments in the data, however, the qualitative features presented in the work should still hold. We shall explore such related issues in a future work.

\section*{Acknowledgements}

BB and TRS acknowledge the facilities at the IRC, DU. TRS acknowledges the R \& D grant from Delhi University. 

\bibliography{detailedfrac_revised_v1,extra}
\bibliographystyle{mnras}
\end{document}